\documentclass[twocolumn,showpacs,showkeys]{revtex4}
\usepackage{graphicx,subfigure,epsfig}
\input{psfig.sty}

\newcommand{\bastar}{\begin{eqnarray*}}
\newcommand{\eastar}{\end{eqnarray*}}
\newskip\humongous \humongous=0pt plus 1000pt minus 1000pt

\newif\ifdtup

\relax
\newcommand{\bea}{\begin{eqnarray}}
\newcommand{\eea}{\end{eqnarray}}
\newcommand{\nn}{\nonumber}

\newcommand{\dfrac}{\displaystyle\frac}
\newcommand{\mn}{{\mu\nu}}

\begin{document}
\title {Hierarchy Problem, Dilatonic Fifth Force, and Origin of Mass}
\author{{Y. M. Cho}$^{}$\footnote{E-mail:~\texttt{ymcho@yongmin.snu.ac.kr}}
and {J. H. Kim}$^{}$\footnote{E-mail:~\texttt{rtiger@phya.snu.ac.kr}}}
\affiliation{Center for Theoretical Physics
and School of Physics  \\ 
College of Natural Sciences,
Seoul National University, Seoul 151-742, Korea  \\}
\author{{S. W. Kim}$^{}$\footnote{E-mail:~\texttt{sungwon@ewha.ac.kr}}}
\affiliation{Department of Science Eudcation,
Ewha Womans University, Seoul 120-750, Korea}
\author{{J. H. Yoon}$^{}$\footnote{E-mail:~\texttt{yoonjh@konkuk.ac.kr}}}
\affiliation{Department of Physics,
Konkuk University, Seoul 143-701, Korea}

\begin{abstract}
Based on the fact that the scalar curvature of the internal
space determines the mass of the dilaton
in higher-dimensional unified theories,
we show how the dilaton mass can explain the origin of
mass and resolve the hierarchy problem. Moreover, we show
that cosmology puts a strong constraint on dilaton mass,
and requires the scale of the internal space to be 
larger than $10^{-9}~m$.
\end{abstract}
\maketitle

One of the fundamental problems in physics is the so called
hierarchy problem. It has been very difficult
to understand why the Planck mass fixed by
the Newton's constant is so large compared to the mass scale of
ordinary elementary particles, or equivalently
why the gravitational force is so weak compared to other forces.
There have been many proposals to resolve this problem.
Long time ago Dirac conjectured that the Newton's constant
may not be a constant but actually a time-dependent
parameter to resolve the problem \cite{dirac,jor}.
This is an attractive proposal, because this conjecture
can naturally be implemented in realistic higher-dimensional
unified theories \cite{jmp75,prd87}.
Another interesting proposal based on the higher-dimensional
unification is that the gravitational force
in higher-dimension is actually as strong as
other forces, but a relatively large (compared to the Planck scale)
internal space of the order of $TeV$ scale makes
the 4-dimensional gravitational force very
weak \cite{ark,dim}.

A totally independent on the surface but actually intimately
related problem is the dilatonic fifth force and
the origin of dilaton mass \cite{prd90,grg91}.
All existing higher-dimensional unified theories (Kaluza-Klein
theory, supergravity, and superstring) predict
the existence of the dilaton, the fundamental scalar graviton
which couples to all matter fields \cite{jmp75,duff,witt}.
As the scalar graviton it generates a fifth force which
can affect the Einstein's gravity in a fundamental way,
and has a deep impact in cosmology.
The reason why the dilatonic fifth force is related to
the hierarchy problem is that the dilaton mass which determines
the range of the fifth force is given by the scalar
curvature of the internal space, whose scale is fixed by
the scale of the internal space \cite{prd87,prd90}.
{\it The purpose of this Letter is to show how the dilaton mass
can resolve the hierarchy problem and allow us
to understand the origin of mass, and to discuss how 
the cosmology excludes the dilaton with mass heavier than 
$160~{\rm eV}$.} This, together with the known
fifth force experiment constraint on dilaton mass, 
requires the scale of internal space 
to be smaller than $10^{-5}~m$, but larger than $10^{-9}~m$.

It has been well known that in higher-dimensional unified theories
the volume element of the internal space has important roles
in 4-dimensional physics \cite{jmp75,prd87}. It
becomes a fundamental scalar graviton known as the Kaluza-Klein
dilaton, which makes the Newton's constant space-time
dependent and naturally realizes the Dirac's
conjecture \cite{prd87,prd90}.
As an essential part of the $(4+n)$-dimensional metric
the dilaton (also called the ``radion" \cite{dim} or
the ``chameleon" \cite{kho}) couples to all matter fields,
and generates the dilatonic fifth force which could compromise
the equivalence principle \cite{grg91,prl92}.
It acquires a mass geometrically, and thus can easily
accommodate the experimental fact that there is no long range
fifth force in nature \cite{exp1,exp2}.
Moreover, in cosmology the dilaton could play the role
of the inflaton, and well be the dark matter
of the universe \cite{prl92,cqg98}.
But as the dark matter it can not have
arbitrary mass. This severely constrains the dilaton mass,
and thus the size of the internal space.
In the following we discuss how the dilaton acquires a mass
and how the existing constraints
on the dilaton mass restrict the size of the internal space.

To do this, it is worth comparing
two dimensional reduction methods in higher-dimensional
theories, because the way the dilaton couples to matter
fields crucially depends on the dimensional reduction methods.
In the popular dimensional reduction one
treats the $(4+n)$-dimensional space as physical and
view the $4$-dimensional physics as an approximation
in the limit when the $n$-dimensional space is very
small \cite{duff,witt}.
On the other hand, in the dimensional reduction by
isometry, one assumes an exact isometry
in the $(4+n)$-dimensional space to make the $n$-dimensional
space unphysical \cite{jmp75,prd87}. In this view
the $n$-dimensional space becomes invisible not because of
the small size but because of the isometry, so that
the size of the $n$-dimensional space can be arbitrary
in principle. It has been thought that the popular view is
more general because it practically reduces to the other view
in the zero mode approximation (where one excludes the massive
Kaluza-Klein modes). However, there are subtle but important
differences between the two reduction methods
which can change the 4-dimensional physics drastically,
as we will see in the following.

Since all higher-dimensional unified theories contain
the $(4+n)$-dimensional gravity we start from the dimensional
reduction of Kaluza-Klein theory. Consider the popular
dimensional reduction first. In the zero mode
approximation, one has to exclude the massive
Kaluza-Klein modes.
A simple way to do this is to introduce an $n$-dimensional
isometry G, and to view the $(4+n)$-dimensional
unified space as a principal fiber bundle
P(M,G) made of the $4$-dimensional space-time M as the base manifold 
and the $n$-dimensional group manifold G as the vertical fiber
(the internal space) on which G acts as an isometry.
Let $\gamma_{\mu\nu}$ and ${\phi}_{ab}$ be
the metric on M and G, $\gamma$ and $\phi$ be the determinants
of $\gamma_{\mu\nu}$ and ${\phi}_{ab}$, and
$\rho_{ab}={\phi}^{-1/n}~{\phi}_{ab}$
($|\textrm{det}~\rho_{ab}|=1$) be the normalized internal metric.
In this setting the (4+n)-dimensional
Einstein-Hilbert action on P leads to the following 4-dimensional
Lagrangian in the Jordan frame \cite{jor,jmp75}
\bea
&\mathcal{L}_0= -\dfrac{\hat V_G}{16\pi G_P}
\sqrt{\gamma}\sqrt{\phi}~\Big[~R_M  \nn\\
&-\dfrac{n-1}{4n}\gamma^{\mu\nu}
\dfrac{(\partial_{\mu}\phi)(\partial_{\nu}\phi)}{\phi}^{2}
+\dfrac{\kappa^2}{4} \root n\of{\phi}
~\rho_{ab}\gamma^{\mu\alpha}\gamma^{\nu\beta}
F_{\mu\nu}^{a}F_{\alpha\beta}^b \nn\\
&+\dfrac{\gamma^{\mu\nu}}{4}(D_\mu \rho^{ab})(D_\nu \rho_{ab})
+\dfrac{1}{\kappa^2 \root n\of{\phi}}~\hat R_G(\rho_{ab}) \nn\\
&+\Lambda_P+\lambda(|\textrm{det}\rho_{ab}|-1)~\Big],
\label{cflag1}
\eea
where $G_P$ is the (4+n)-dimensional gravitational constant,
$\hat V_G$ is the normalized volume of the internal space G,
$R_M$ is the scalar curvature of M fixed by $\gamma_{\mu\nu}$,
$\hat R_G(\rho_{ab})$ is the normalized
internal curvature of G fixed by $\rho_{ab}$,
\bea
&\hat R_G(\rho_{ab})=-\dfrac12 f_{ab}^{~~d}f_{cd}^{~~b} \rho^{ac}
-\dfrac 14 f_{ab}^{~~m}f_{cd}^{~~n}\rho^{ac}\rho^{bd}\rho_{mn}, \nn
\eea
$\kappa$ is the unit scale of the internal space,
$F_{\mu\nu}^{a}$ is the gauge field of the isometry group,
$\Lambda_P$ is the (4+n)-dimensional cosmological constant,
and $\lambda$ is a Lagrange multiplier.
To clarify the meaning of (\ref{cflag1}), we introduce the Pauli
metric (also called the Einstein metric in string theory)
$g_{\mu\nu}$ and the Kaluza-Klein dilaton $\sigma$ by
\bea
&g_{\mu\nu}=\exp \big(\sqrt{\dfrac{n}{n+2}}~\sigma \big)
\gamma_{\mu\nu}, \nn\\
&\phi= \Big[v~\exp \big(\sqrt{\dfrac{n}{n+2}}~\sigma \big)
\Big]^2,~~~<\phi>=v^2.
\eea
With this (\ref{cflag1}) is expressed in
the Pauli frame as \cite{jmp75,prl92}
\bea
&\mathcal{L}=-\dfrac{v~\hat V_G}{16\pi G_P}~\sqrt{g}
\Big[R+\frac{1}{2}(\partial_{\mu}\sigma)^2
-\dfrac{1}{4}(D_\mu \rho^{ab})(D^{\mu}{\rho}_{ab}) \nn\\
&+\dfrac{1}{\kappa^2}~v^{-2/n}~\hat R_G(\rho_{ab})
\exp \big(-\sqrt{\dfrac{n+2}{n}}~\sigma \big) \nn\\
&+\Lambda_P \exp \big(-\sqrt{\dfrac{n}{n+2}}~\sigma \big) \nn\\
&+\lambda\exp \big(-\sqrt{\dfrac{n}{n+2}}\sigma\big)
(|\textrm{det}\rho_{ab}|-1) \nn\\
&+\dfrac{\kappa^2}{4}~v^{2/n}~\exp
\big(\sqrt{\dfrac{n+2}{n}}~\sigma \big)
\rho_{ab} {F}_{\mu\nu}^{a} {F}^{b \mu\nu}~\Big].
\label{cflag}
\eea
This describes the well-known unification of gravitation
with the gauge field, if one identifies \cite{jmp75,prd87}
\bea
&\dfrac{v~\hat V_G}{16\pi G_P}=\dfrac{1}{16\pi G},
~~~~~\dfrac{\kappa^2}{16\pi G}=1,
\label{newc}
\eea
where $G$ is the Newton's constant.

This tells the followings.
First, although the unit scale of
the internal space $\kappa$ is fixed by the Planck scale
$\sqrt{16\pi G}$, the vacuum
expectation value of the volume of the internal space is
given by \cite{prd90,grg91}
\bea
&<V_G>=v~\hat V_G
\simeq v~\kappa^n=v~(16\pi G)^{n/2}.
\eea
This means that the actual scale of the internal space $L_G$
is given by $v^{1/n} \kappa$, not the Planck scale.
Perhaps more importantly, this tells that the scale of
the higher-dimensional gravitational constant
$G_P$ is given by
\bea
&{G_P}^{1/(n+2)}=(16\pi)^{n/2(n+2)}~{v}^{1/(n+2)}~G^{1/2}.
\label{veq}
\eea
This is precisely the equation which has been proposed
to resolve the hierarchy problem \cite{ark,dim},
which shows that a large $v$ can easily bring $G_P$
to the order of the elementary particle scale.
Of course, in the popular dimensional reduction
in which the $(4+n)$-dimensional space is treated as physical,
the internal space can not assume a large scale
because it has to be invisible at present
energy scale. For this reason the size has often been assumed to be
of the Planck scale, with $v=1$ \cite{jmp75,duff,witt}.
But we emphasize that a relatively large internal space has not been
ruled out theoretically as well as experimentally \cite{prd90,exp1}.

Second, it is the Pauli metric $g_\mn$, not the Jordan metric
$\gamma_\mn$, which describes the massless spin-two graviton
of Einstein's theory \cite{prl92}.
But notice that (in the absence of the dilaton)
its coupling to the gauge field depends on $v$,
so that the graviton couples to the gauge field non-minimally
when $v\neq 1$. And this leads to a violation of
the equivalence principle \cite{grg91,prl92}. Fortunately
we can remove this $v$-dependent gravitational coupling
by renormalizing $F_{\mu\nu}^{a}$ to
${\hat F}_{\mu\nu}^{a}=v^{1/n} F_{\mu\nu}^{a}$ and identifying
${\hat F}_{\mu\nu}^{a}$ as the $4$-dimensional gauge field.
When one has higher-dimensional matter fields (as is the case
in superstring or supergravity), however,
the rescaling of the matter fields to absorb
the $v$-dependent gravitational coupling
is no longer possible in the popular dimensional reduction.
This is because the normalization
of the matter fields is pre-fixed by the $(4+n)$-dimensional
physics, since the higher-dimensional space is treated as
physical here \cite{ark,dim}.
In this case a large $v$ inevitably leads to a disastrous
violation of the equivalence principle in Einstein's theory,
which implies that the scale of the internal space can not
be much larger than the Planck scale.
This is a potentially dangerous defect of the popular dimensional
reduction which has to be treated very carefully.

However, one can avoid this violation of the equivalence
principle if one adopt the dimensional reduction by isometry.
This is because in this case one can renormalize
the matter fields and their masses with impunity after
the dimensional reduction to absorb
the $v$-dependent gravitational coupling, since only
the 4-dimensional space-time is treated as physical in
this dimensional reduction. In this case one can have a large
$v$ without violating the equivalence principle.
This tells that the dimensional reduction by isometry
has a logical advantage over the popular
dimensional reduction \cite{jmp75,pen}.

Suppose the Lagrangian (\ref{cflag}) has the unique vacuum at
$<g_{\mu\nu}>=\eta_{\mu\nu},~<\sigma>=0,
~<\rho_{ab}>=\delta_{ab},~<A_\mu^a>=0$.
With this we have the following dilaton potential
$V(\sigma)$,
\bea
&V(\sigma)=v^{-2/n}~\dfrac{\hat R_G}{(16\pi G)^2}
~\Big[\exp \big(-\sqrt{\dfrac{n+2}{n}}~\sigma \big) \nn\\
&-\dfrac{n+2}{n}~\exp \big(-\sqrt{\dfrac{n}{n+2}}~\sigma \big)
+\dfrac2n \Big],
\label{dpot}
\eea
where $\hat R_G=\hat R_G(\delta_{ab})$ is the (dimensionless)
curvature of G created by the Cartan-Killing metric $\delta_{ab}$.
Notice that the potential is completely fixed by
the vacuum condition $V(0)=0$ and $dV(0)/d\sigma=0$.
Clearly $V(0)=0$ assures that we have no {\it ad hoc}
$4$-dimensional cosmological constant,
and $dV(0)/d\sigma=0$ shows that the $(4+n)$-dimensional
cosmological constant $\Lambda_P$ is uniquely fixed.
From (\ref{dpot}) we find the mass of the Kaluza-Klein
dilaton $\mu$,
\bea
&\mu^2
=-v^{-2/n}~\dfrac{\hat R_G}{8\pi n}~m_p^2
=\dfrac2{n+2} \Lambda_P,
\label{dmass}
\eea
where $m_p$ is the Planck mass.
This demonstrates that a large $v$ naturally transforms
the large Planck mass to a small dilaton mass, which can easily
be of the order of the elementary particle mass scale.
This is the resolution of the hierarchy problem
in Kaluza-Klein unification \cite{prd87,prd90}.
But notice that $\mu=0$ when
$\hat R_G=0$, independent of $n$ and $v$.
From (\ref{dmass}) we have
the following scale of the internal space $L_G$
(when $\hat R_G \neq 0$),
\bea
L_G=v^{1/n}~\kappa=\sqrt{-\dfrac{2\hat R_G}n}
~\dfrac1{\mu} \simeq \dfrac1{\mu}.
\label{ints}
\eea
So, for the $S^3$-compactification of $3$-dimensional internal
space in $(4+3)$-dimensional unification with G=SU(2),
we have $\hat R_G=-3/2$ and $L_G=1/\mu$.

At this point it is worth comparing (\ref{veq}) and (\ref{dmass}).
Both provide a resolution of the
hierarchy problem. But there is a big difference. Clearly
(\ref{veq}) does that making the higher-dimensional Newton's constant
large, but (\ref{dmass}) does that with the dilaton mass.
Moreover, the dimension of the internal space $n$ plays the crucial
role in (\ref{veq}). But the curvature of the internal space 
plays the crucial role in (\ref{dmass}). In fact here 
it is crucial that we have a non-vanishing $\hat R_G$ 
to resolve the hierarchy problem. 
Furthermore, (\ref{dmass}) solves the hierarchy problem 
providing a new mass generation mechanism,
a geometric mass generation through the curvature of space-time,
which tells that the hierarchy problem is closely related to 
the problem of the origin of mass. More significantly, 
it tells that the curvature can be the cause (not the effect) 
of the mass. Understanding the origin of mass 
has been a fundamental problem in physics. 
The geometric mass generation mechanism could provide a
natural resolution to this problem.

As the scalar graviton the dilaton modifies Einstein's
gravitation in a fundamental way \cite{grg91,prl92}.
To see how, notice that the sum of the gravitational
and fifth force between the two baryonic point particles
separated by a distance $r$ is given in the Newtonian limit as
\bea
F\simeq\frac{\alpha_{g}}{r^{2}}
+\dfrac{\alpha_{5}}{r^{2}} e^{-\mu r}
=\dfrac{\alpha_{g}}{r^{2}} \big(1+\beta e^{-\mu r} \big),
\eea
where $\alpha_g$and $\alpha_5$ are the fine structure constants of
the gravitation and fifth force, and $\beta$ is the ratio between
them. In terms of Feynman diagrams the first term represents one
graviton exchange but the second term represents one dilaton
exchange in the zero momentum transfer limit. In Kaluza-Klein
unification we have $\beta=n/(n+2)$ \cite{prd87,prd90},
but in general one may assume $\beta \simeq 1$ because
the dilaton is the scalar partner of
the graviton. With this assumption one may try to measure
the range of the fifth force experimentally.
A recent torsion-balance fifth force experiment puts
the upper bound of the range to be around $44~\mu m$
(and the dilaton mass to be around $4.5\times 10^{-3}~eV$)
with 95\% confidence level \cite{exp2}. This, with (\ref{ints}),
implies that in the $(4+3)$-dimensional unification
with G=SU(2) the size of the internal space $L_G$
can not be larger than $44~\mu m$.

On the other hand the cosmology puts a strong theoretical
constraint on dilaton mass, because the dilaton can easily be 
the dominant matter of the universe. In cosmology 
the dilaton starts with the thermal equilibrium 
at the beginning and decouples from other sources 
very early near the Planck time. But it is not stable,
although it will decay very slowly due to the weak (i.e.,
the gravitational) coupling to the matter fields.
There are two dominant decay channels of the dilaton,
two-photon decay and fermion-antifermion decay,
which may be described by the following Lagrangian \cite{cqg98},
\bea
&\mathcal{L}_{int} \simeq -\dfrac{1}{4} g_{1} \sqrt{16 \pi G}
~\hat{\sigma}~F_{\mu\nu}F^{\mu\nu} \nn\\
&-g_{2} \sqrt{16 \pi G}~m~\hat{\sigma}~\bar{\psi}\psi,
\label{ddint}
\eea
where $g_1$ and $g_2$ are dimensionless coupling constants,
$m$ is the mass of the fermion, and
$\hat \sigma=\sigma/\sqrt{16\pi G}$
is the dimensional (physical) dilaton field.
From this we can calculate the two-photon decay rate and
the fermion-antifermion decay rate of
dilaton in tree approximation \cite{cqg98}
\bea
&\Gamma_{\sigma\rightarrow\gamma\gamma}
=\dfrac{g_1^2 \mu^3}{16~m_p^2},  \nn\\
&\Gamma_{\sigma\rightarrow\bar{\psi}\psi}
= \dfrac{2 g_2^2~m^2~\mu}{m_p^2}\times
\Bigg[1-\bigg(\dfrac{2m}{\mu}\bigg)^2\Bigg]^{3/2}.
\label{ddr}
\eea
With this we can estimate the present number density
of the relic dilaton $n(t_{0})$,
\bea 
&n(t_{0}) \simeq 7.5~\exp \Big(\dfrac{-t_0}{\tau(\mu)} \Big)
~\textrm{cm}^{-3},
\eea
where $t_0=1.5\times10^{10}~\textrm{yr}$ is the age of
the universe and $\tau(\mu)$ is
the total life-time of the dilaton. 
This implies that the dilaton can easily survive to 
the present universe and can be the dominant matter of
the universe 

Now, for the dilaton to be the dark matter, we must have
the following constraints for the dilaton mass.
First, the dilaton mass density $\rho(\mu)$ must be equal to
the dark matter density $\rho_0 (\mu)$ \cite{cqg98},
\bea
&\rho(\mu)=\mu \times 7.5 ~\exp \Big(\dfrac{-t_0}{\tau(\mu)} \Big)
~\textrm{~cm}^{-3}  \nn\\
&\simeq \rho_0 (\mu) \simeq 0.23\times 5.9~\textrm{keV cm}^{-3}.
\label{dmass1}
\eea
Secondly, the energy density of the daughter particles
(photons and light fermions) $\tilde \rho(\mu)$ must be
negligibly small compared to the dark matter density,
\bea
\tilde \rho(\mu) \ll \rho_0 (\mu).
\label{dmass2}
\eea
To find the dilaton mass which satisfies the above constraints,
we have to know the coupling constants
$g_1$ and $g_2$.  In Kaluza-Klein unification they are
given by \cite{prd87}
\bea
&g_1=\sqrt{\dfrac{n+2}{n}},~~~~g_2=\sqrt{\dfrac{n}{n+2}}.
\eea
But here we leave them as free parameters,
assuming only $g_1\simeq g_2$.

The first constraint (\ref{dmass1}) shows that, 
when $g_1\simeq g_2 \simeq 1$, there are 
two mass ranges of dilaton in which the relic dilaton 
could be the dominant matter of the universe;
$\mu \simeq 160~eV$ with life-time
$\tau\simeq 3.8 \times 10^{35}~sec$
and $\mu\simeq 276~MeV$ with life-time
$\tau\simeq 3.3 \times 10^{16}~sec$. The dilaton with mass smaller than
$160~eV$ survives but fails to be dominant matter due to its low mass,
and the dilaton with mass larger than $276~MeV$ does not
survive long enough to become the dominant matter of the universe.
The dilaton with mass in between is impossible
because it would overclose the universe. This clearly rules out 
the ADD dilaton with mass of $TeV$ range. 

Moreover, the second constraint (\ref{dmass2}) shows that 
the dilaton with mass $276~MeV$ can not be aceptable as 
the dark matter. To see this notice that 
$\tau_2 \simeq 6.9 \times 10^{-2}~t_0$, so that most of 
the dilaton with mass $276~MeV$ have already decayed. 
Indeed only $1.98 \times 10^{-6}$ of the heavy
dilatons which survive now fill the dark matter energy. 
This means that the energy density of the daughter particles from
the heavy dilaton must be much bigger than the energy density 
of the dilaton itself, which tells that the daughter particles from
the heavy dilaton overclose the universe. This makes 
the heavy dilaton unacceptable. On the other hand 
the dilaton with mass $160~eV$ is almost stable 
because $\tau_1 \simeq 8.1 \times 10^{17}~t_0$.
In this case the energy density of the daughter particles from
the light dilaton is negligible compared to the energy density 
of the dilaton itself, so that the $160~eV$ dilaton can easily 
satisfy the second constraint. In conclusion only 
the $160~eV$ dilaton can become the dark matter of the universe. 
This puts a strong cosmological constraint on dilaton mass,
and immediately rules out the internal space of the scale 
smaller than $1.2~nm$.

\begin{figure}
\epsfig{file=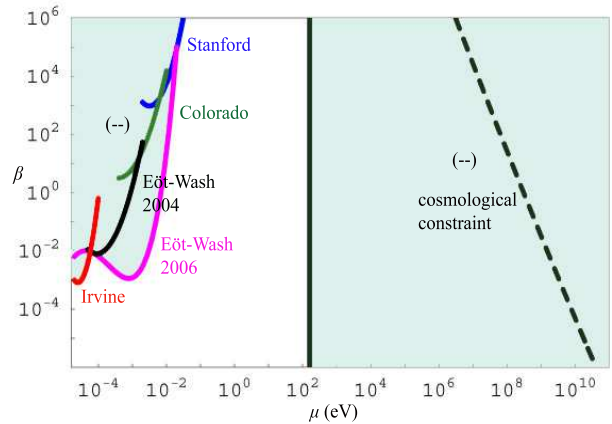, height = 6cm, width = 8.5cm}
\caption{\label{Fig. 1} The allowed mass $\mu$
of dilaton (uncolored region), where we leave $\beta=\alpha_5/\alpha_g$
arbitrary. The colored region marked by (--)
is the excluded region, and the dotted line represents 
the mass of the heavy dilaton whose daughter particles 
overclose the universe.}
\end{figure}

Putting these constraints together we obtain Fig. 1, 
which shows the allowed region of the dilaton mass
(and the allowed range of the fifth force)
in $7$-dimensional unification with $S^3$ compactification
of the internal space. Although $\beta$
is expected to be of the order one, we leave it arbitrary here.
Here we emphasize that the cosmological estimate of the dilaton mass 
discussed in this paper should be understood as an order estimate, 
not an exact result, because it is based on the linear approximation.
Nevertheless our result demonstrates that cosmology provides
a crucial piece of information on the dilaton mass. 
In particular, cosmology rules out the $TeV$ scale internal space 
suggested by ADD. 
Moreover, our result implies that the dilatonic
fifth force may be too short-ranged to be detected by
the fifth force experiment. Under this circumstance
a totally different type of experiments based on two-photon
decay of dilaton is needed to detect the dilaton \cite{cho07}.

In this paper we have discussed how the dilaton mass can 
resolve the hierarchy problem and explain the origin of mass.
Moreover, we have shown how the dilaton can be the dominant 
matter of the universe, and how the cosmology can determine 
the dilaton mass the scale of the internal space. 
In particular, we have shown that the cosmology effectively 
excludes the dilaton heavier than $160~{eV}$, and thus 
the internal space whose scale is smaller than $10^{-9}~m$. 

We close with the following remarks. \\
1. So far we have concentrated on the dilaton. But we emphasize
that there are other scalar gravitons called internal gravitons
(also called the moduli in string theory)
which are described by $\rho_{ab}$, which have similar
properties \cite{jmp75,prd87}. Here we simply mention
that in general there are $(n+2)(n-1)/2$
such internal gravitons which have similar features
and thus enhance the impact
of the dilaton we have discussed above. \\
2. In superstring or supergravity the situation is more complicated.
For example, in string theory one has to deal with 
an extra higher-dimensional dilaton
(the string dilaton) which remains massless in
all orders of perturbation \cite{witt}.
But once the dilaton acquires a mass, the qualitative
features of dilaton physics will remain the same.
In particular, the constraint on the dilaton mass 
shown in Fig.1 must apply to all higher-dimensional 
unified theories. \\
3. In the above analysis, we have assumed that
$<\sigma>=0$. But in cosmology the dilaton field
(as the inflaton) may actually evolve in time,
so that classically $\sigma(t_0)$
may not be zero at present time $t_0$ \cite{prd90}.
Even in this case our conclusion may still be valid,
if we identify $v$ and $\hat R_G$ as the present volume
and curvature of the internal space
(with $\sigma(t_0)=0$) \cite{prd90,prl92}.

The detailed discussions of our results and related subjects
will be published in a separate paper \cite{cho07}.

{\bf ACKNOWLEDGEMENT}

~~~The work is supported in part by the BSR Program (Grant
KRF-2007-314-C00055) of Korea Research Foundation  
and by the International Cooperation Program and 
ABRL Program (Grant R14-2003-012-01002-0) of 
Korea Science and Engineering Foundation.

\end{document}